\PassOptionsToPackage{usenames,dvipsnames,table}{xcolor}
\documentclass[twocolumn,aps,prd,longbibliography]{revtex4-1}
\usepackage{graphicx,epsfig,epstopdf}
\usepackage{amsmath,amssymb,latexsym,bm}
\usepackage[english]{babel}
\usepackage{wrapfig}
\usepackage{animate}
\usepackage{times}
\usepackage[T1]{fontenc}
\usepackage{color, colortbl}
\usepackage{tikz}
\usepackage[usenames, dvipsnames]{xcolor}
\usetikzlibrary{arrows,shapes}
\usepackage{url}
\usepackage{hyphenat}
\usepackage{makecell}
\usepackage[Conny]{fncychap}
\usepackage{lipsum}
\usepackage{mathptmx}
\usepackage{mathtools}
\usepackage[T1]{fontenc}
\usepackage[%
  colorlinks=true,
  urlcolor=blue,
  linkcolor=blue,
  citecolor=blue
]{hyperref}
\DeclareUnicodeCharacter{2212}{-}
\newcommand{\orcid}[1]{\href{https://orcid.org/#1}{\resizebox{10px}{!}{\includegraphics{orcid.png}}}}
\begin{document}
\title{Seeing dark matter via acceleration radiation}

\author{Syed Masood A.~S. Bukhari$^{1}$} 
\author{Li-Gang Wang$^{1}$} \email{lgwang@zju.edu.cn}
\affiliation{$^{1}$ School of Physics, Zhejiang University, Hangzhou 310027, China.}
\begin{abstract}
 Despite constituting a noteworthy $\sim 27\%$ share of the total energy budget of our Universe, dark matter (DM) has thus far eluded direct observations. Owing to its pervasive nature, there is a sincere expectation that astrophysical black holes (BHs) encompassed by DM should leave distinctive imprints on the gravitational waves arising from BH mergers. Theoretical models of DM present a diverse landscape of possibilities, with perfect fluid dark matter (PFDM) emerging as a recent and notably intriguing candidate model. In this work, utilizing the established quantum optical approach, we investigate the possibility of catching DM signatures via acceleration radiation emitted by a freely-falling detector (e.g. an atom) within a PFDM-surrounded Schwarzschild BH. The setup involves a Casimir-type apparatus where the detector interacts with the field, and this situation induces excitations in the detector in a manner consistent with Unruh effect. We observe that our DM candidate, while making classical contributions to spacetime geometry, has the potential to leave quantum imprints in the radiation flux. Notably, it is observed that, in comparison to a pure Schwarzschild BH, PFDM can markedly reduce particle emission as long as its density remains below a critical threshold, and vice versa. Given the lessons we have learnt from realizing cosmological phenomena in simulated laboratory conditions, there is a remote possibility that such study may perhaps provide insights (to whatever degree !) into the future table-top experiments in analogue gravity paradigm.   
    \end{abstract}
\date{\today}
\maketitle

\section{Introduction}
\begin{figure}[tbhp]
\centering
\includegraphics[width=1.0\linewidth, height=8.5cm]{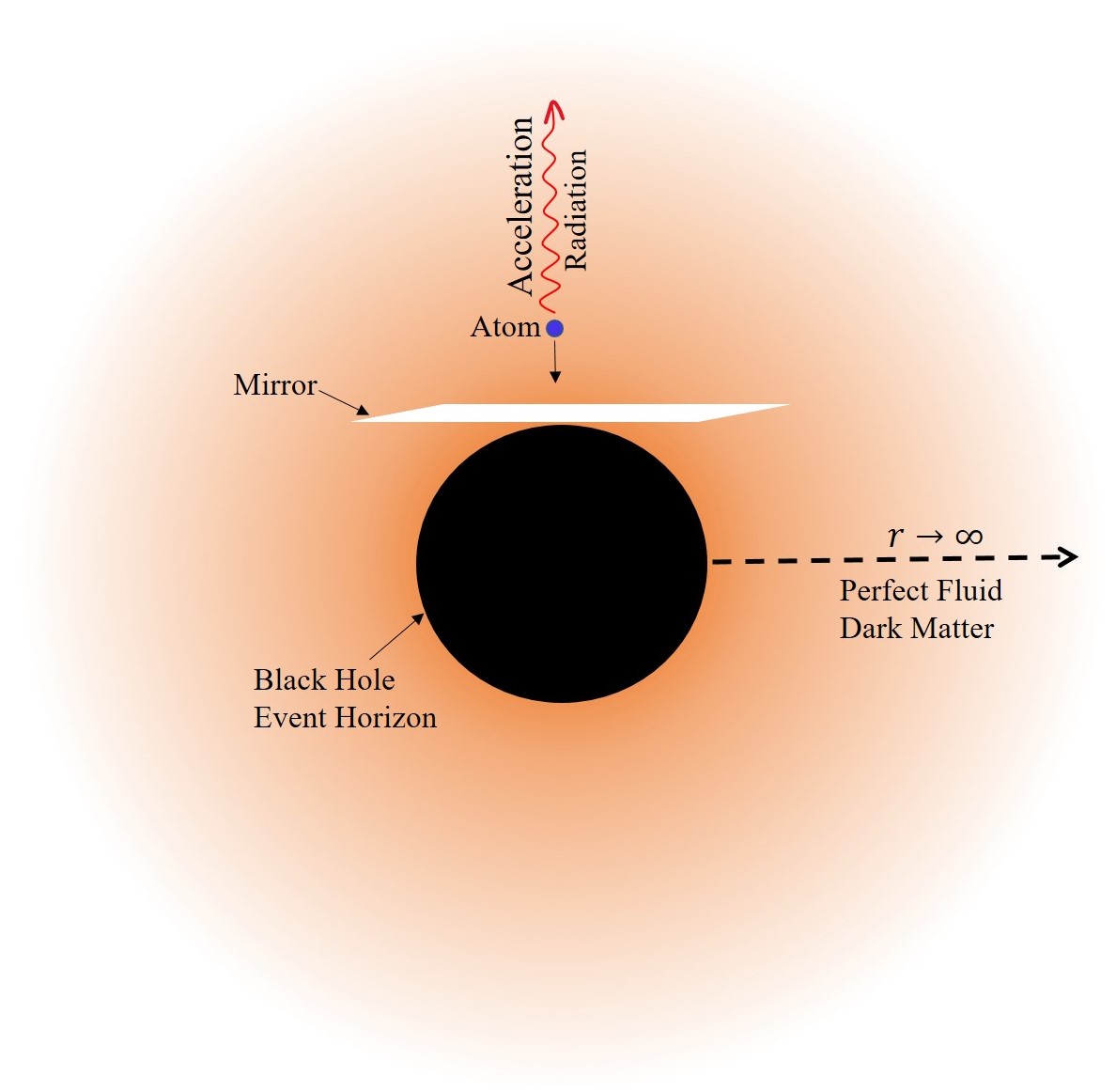}
\caption{Schematic view  of acceleration radiation emitted from a two-level atom falling into a BH in presence of PFDM halo.}
\label{schematic}%
\end{figure}

Hawking radiation \cite{1975CMaPh..43..199H} and  Unruh effect \cite{Unruh:1976db} are the profound insights from  quantum field theory in curved spacetime (QFTCS) \cite{Birrell:1982ix,2009qftc.book.....P}. Although, in both of these phenomena, field quanta are created out of vacuum,  they however differ in the underlying spacetime geometry. Hawking radiation assumes curved background of black holes (BHs), while as Unruh effect is registered  by  accelerated (Rindler) observers in flat Minkowski spacetime \cite{RevModPhys.80.787}.   
These processes arise out of the same principle to that of  \textit{dynamical Casimir  effect} of Moore \cite{1970JMP....11.2679M},  where accelerated boundaries (mirrors) hammer the vacuum to radiate real particles \cite{Svidzinsky:2018jkp}. One more example that shares similar physics is the cosmological particle emission due to expanding spacetimes \cite{2009qftc.book.....P}. For this interesting overlap of atoms, fields and geometries, we refer the reader to our recent review article \cite{Bukhari:2023yjt}. \\
\indent Our Universe does not only contain the familiar baryonic matter, it is rather believed to be filled with all kinds of mysterious dark stuff: dark matter (DM) and dark energy. In fact, there is a mounting indirect evidence for the existence of DM in our Universe, and  current estimates put it at approximately $27\%$ of the total matter-energy content of the Universe \cite{Arbey:2021gdg}. It is therefore more realistic to expect the influence of this \textit{`exotic'} matter on those vacuum-originating  phenomena.\\
 \indent From a cosmological scenario, DM is believed to be underpinning some key phenomena, such as galactic dynamics \cite{Sadeghian:2013laa}.   Since at the moment, we do not have direct experimental/observational evidence of DM, there has been  a flurry of excitement for different models. Some notable candidates include cold dark matter (CDM) \cite{Navarro:1995iw,Navarro:1996gj}, self-interacting DM \cite{Spergel:1999mh}, Bose-Einstein condensation  DM \cite{Hu:2000ke}, superfluid DM \cite{Berezhiani:2015bqa}, and primordial black hole DM \cite{Carr:2021bzv,Escriva:2022duf}, each with its own potential set of working assumptions.   Even though we have this long list of possibilities, however,  the doors for new candidate models have not been shut as DM is still an outstanding problem in modern cosmology and astrophysics. Some time ago, Kiselev \cite{Kiselev:2003ah,Kiselev:2004vy} and others\cite{Li:2012zx} proposed what is now known as perfect fluid dark matter (PFDM) to account for asymptotic rotational curves of spiral galaxies. Since then, this DM candidate has been scrutinized in various scenarios, like BH shadow \cite{Hou:2018avu}, thermodynamics\cite{Zhang:2020mxi}, particle dynamics \cite{Li:2022dwd}, accretion disks \cite{Pugliese:2022oes}, quasinormal modes (QNMs) \cite{Jusufi:2019ltj},  and more. The interesting aspect of QNMs is that they find connections to gravitational waves originating from coalescing BHs \cite{Barack:2018yly}.\\ 
 \indent In this \textcolor{blue}{work}, we undertake a novel route whereby PFDM might manifest via atom-field interactions. Namely, we consider the emission of acceleration radiation by a freely-falling atom in a Schwarzschild BH surrounded by PFDM halo. There is a Casimir boundary (mirror) sitting close to BH event horizon which is the source of accelerated field modes with which the falling atom interacts in line with the Unruh's predictions \cite{Unruh:1976db} . A schematic view of our setup is shown in Fig.\ref{schematic} (see discussion in Sec.\ref{geom} A). We provide evidence that PFDM has the potential to either degrade or enhance the radiation  intensity. Given the recent advancements made in experimental search for DM \cite{Manley:2020mjq,XENONCollaboration:2023dar}, it seems credible that future cosmological observations or table-top settings in analogue gravity program \cite{Barcelo:2005fc,Braunstein:2023jpo} might perhaps be able to test and constrain the theoretical parameter space of DM. We believe this study, to the best of our knowledge, is the first attempt to incorporate DM in quantum optical or Casimir paradigm. \\
 \indent The paper is organized as follows: In the next section, we lay down the necessary structure for understanding our working principle.  Sec.\ref{secpDM} is devoted to the calculation of excitation probability. Results and  discussions  can be found in Sec.\ref{discussionDM}. We draw conclusions in  Sec.\ref{secconclusions}.

\section{The underlying Geometry and the Klein-Gordon equation}\label{geom}
\subsection{The schematic setup}
We note that our work assembles ideas from Casimir physics, QFTCS and cosmological settings. Hence, for the sake of clarity, it becomes mandatory to sketch  a bird's eye view of the different elements involved in our analysis. To this end, we draw a schematic picture in Fig.\ref{schematic}.  \\
\indent Fig. \ref{schematic} represents a Schwarzschild  BH at the center (black color) with the periphery depicting its event horizon. The Casimir boundary (Mirror), shown in white color, hovers over the BH horizon and is in an accelerated frame of reference as outlined by general relativity (GR). It also veils the atom from any Hawking flux emanating from  the BH.  The PFDM halo, shown in orange color, surrounds the BH in such a way that its density is maximum near the BH horizon and monotonically goes to zero at  $r\rightarrow\infty$, thereby perfectly reproducing Minkowski spacetime. The freely-falling atom emits acceleration radiation (red color) which is received by an observer at asymptotic infinity. The scalar quantum field surrounding the BH is assumed to be in a Boulware vacuum state defined by the asymptotic observer. 
\subsection{Horizon structure}
\begin{figure}[tbhp]
\centering
\includegraphics[width=1.0\linewidth, height=6.5cm]{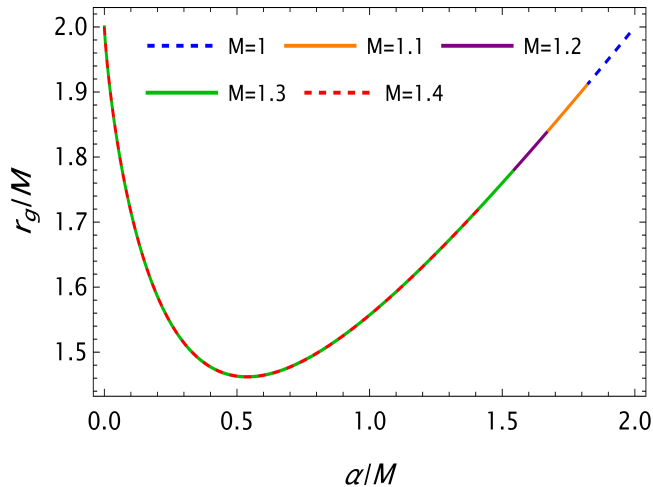}
\caption{Impact of positive PFDM density $\alpha$ on BH horizon radius $r_{g}$ for different masses $M$. Here $\alpha=0$ corresponds to the Schwarzschild BH, for which $r_{g}=2M$.}%
\label{rgalpha}%
\end{figure}

The static, spherically symmetric metric of a BH immersed in a PFDM halo is  given by \cite{Kiselev:2003ah,Kiselev:2004vy,Li:2012zx},
\begin{eqnarray}
ds^{2}=-f(r)dt^{2}+\frac{1}{f(r)}dr^{2}+r^{2}(d\theta ^{2}+\sin^{2}\theta d\phi ^{2}),
\label{metricDM}
\end{eqnarray}
where,
\begin{eqnarray}\label{frDM}
f(r)=1-\frac{2M}{r}+\frac{\alpha}{r}\ln{\left(\frac{r}{|\alpha|}\right)},
\end{eqnarray}
where $\alpha$ is the contribution from PFDM.  The stress energy-momentum tensor of the PFDM distribution is given by 
$T^\mu_\nu={\rm diag}(-\rho,p_{r},p_{\theta},p_{\phi})$,
 where the density, radial and tangential pressures, respectively read as 
$\rho=-p_{r}= \frac{\alpha}{8\pi r^3}$,   $p_{\theta}=p_{\phi}=\frac{\alpha}{16\pi r^3}$.
The value of $\alpha$ can be both positive and negative, and is constrained theoretically as
$0 < \alpha < 2M$ and $−7.18M < \alpha < 0$, respectively  \cite{Xu:2017bpz}. As for $\alpha>0$, it is a direct consequence of the weak energy condition of GR, ensuring  a positive energy density. Although the case $\alpha<0$ finds its mention in few works \cite{Xu:2017bpz,Haroon:2018ryd}, however, to the best of our understanding, its true physical  meaning is still obscure as its represents a negative energy density. Negative energy densities violate the energy conditions of GR and classify  as hypothetical matter distributions.  Therefore, in the present work, we only consider  $\alpha>0$.
For $\alpha=0$, the above metric reduces to that of a Schwarzschild BH.
 By determining roots of $f(r)$, one  gets 
\begin{eqnarray}
    r_{g}=\alpha  W\left[\exp\left(\frac{2 M}{\alpha}\right)\right], 
\end{eqnarray}
which is the BH event horizon. Here  $ W\left[ \cdot \right]$ is the Lambert $W$ function. It is widely accepted that the presence of DM does not alter the number of horizons, and it rather affects the horizon size only. To quantify its effect on the BH, we plot $r_{g}$ in Fig. \ref{rgalpha} against different PFDM density profiles (i.e. $\alpha$).

It can be readily seen from Fig. \ref{rgalpha} that, compared to a pure Schwarzschild BH, the presence of PFDM reduces the BH horizon radius monotonically until a certain minimum value $\alpha_{c}$ is reached, and then increases it again.   This minimum radius corresponds to a certain critical value of $\alpha$, which differentiates the two ways that PFDM alters the geometry of our BH. The situation in the former case $(\alpha<\alpha_{c})$ is similar to that of a charged or  rotating BH (or a BH with some extra `hair') whose radius is usually less compared to the Schwarzschild case. In the  latter case $(\alpha>\alpha_{c})$, PFDM increases $r_{g}$ and depicts an effective mass contribution to the BH, which, from a geometric perspective, is much  like what dark energy does \cite{Bhattacharya:2018ltm}. The critical value of $\alpha$ is gotten by $d r_{g}/d\alpha=0|_{M=const}$ and  comes out to be    
\begin{eqnarray}\label{alphac}
    \alpha_{c}=\frac{2 M}{1+e},
\end{eqnarray}

which seems an interesting number due to appearance of $e$ in the denominator. Note that our plot axes are normalized by $M$ as the gist of the graph is to understand the nature of $r_{g}$ and $\alpha_{c}$. However, from a quantitative perspective, one realizes that for different values of BH mass $M$ in Fig. \ref{rgalpha}, $\alpha_{c} \sim 0.538, 0.592$, $0.645$,$0.70$, and $0.753$ respectively, corresponding to $M=1,1.1,1.2,1.3$, and $1.4$.  To verify whether $r_{g}$ from $f(r)=0$ corresponds to the spacetime singularity or not, one can calculate Kretschman scalar from Riemann tensor $R_{\mu\nu\eta\delta}$, i.e.,  
 $K=R_{\mu\nu\eta\delta}R^{\mu\nu\eta\delta}$, given by \cite{PhysRevD.102.104062}
\begin{eqnarray}\nonumber
K&=&\frac{1}{r^6}\Bigg[12 \alpha^2 \ln^2\left(\frac{r}{ \alpha }\right)+13 \alpha^2-4 \alpha \ln\left(\frac{r}{	\alpha }\right)\\
&\times& \left(12M+5 \alpha\right)+ 40\alpha M+48 M^2\Bigg],
	\end{eqnarray}
 which obviously diverges at $r=0$ only.
 Hence the true physical singularity appears only at $r=0$, the center of the BH, as seen from the above equation.

For this spacetime geometry, the tortoise coordinate reads
\begin{align}
    r_{*}&=\int \frac{dr}{f(r)}\\
    \label{rstar}
    &=\int \frac{dr}{1-\frac{2M}{r}+\frac{\alpha}{r}\ln{\left(\frac{r}{|\alpha|}\right)}},
\end{align}
which is quite difficult to be solved analytically. We will numerically estimate it and use in final probability distribution given in Sec. \ref{secpDM}.

\subsection{Geodesic equations}
\begin{figure*}[t]
\centering
\includegraphics[width=\linewidth, height=13cm]{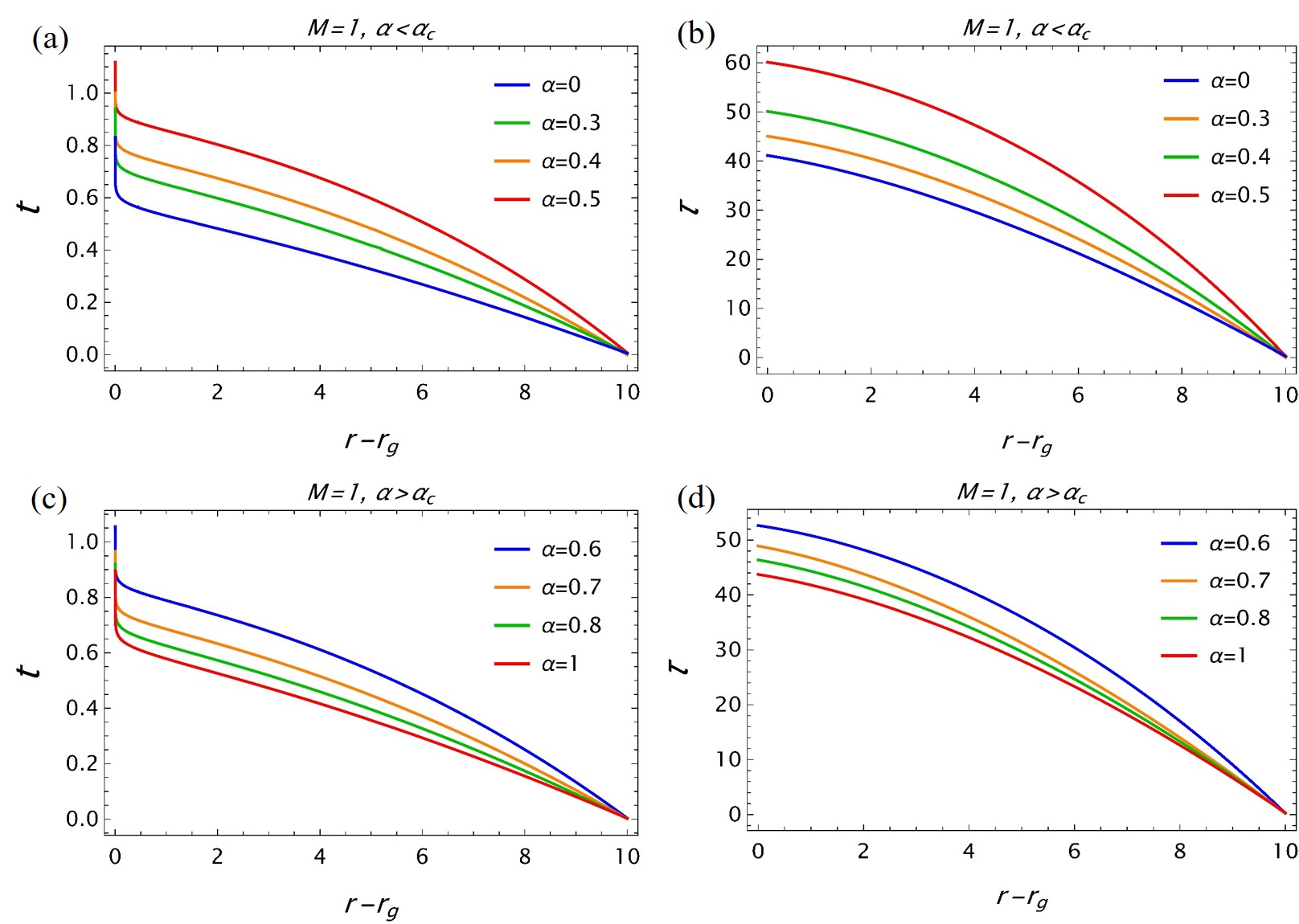}
\caption{Impact of $\alpha$ on the behaviour of coordinate time $t$ and proper time $\tau$ against a rescaled radial coordinate $r-r_{g}$.  (a) and (b), respectively, show $t$ and $\tau$ for the case $\alpha<\alpha_{c}$ respectively, while as (c) and (d) represent quantities for the case $\alpha>\alpha_{c}$.   $\alpha$ increases both of them for  $\alpha<\alpha_{c}$, and vice versa. For $M=1$, $\alpha_{c}\sim 0.538$.}%
\label{timeplotDM}%
\end{figure*}

The solution to the  geodesic equations for a radially infalling atom, following a timelike geodesic, helps us to compute the coordinate time $t$ and proper (conformal) time $\tau$. We thus have \cite{1992mtbh.book.....C}
\begin{eqnarray}
 \frac{d^2 x^{\mu}}{d\tau^2}+\Gamma_{\rho \sigma}^{\mu}\frac{dx^{\rho}}{d\tau}\frac{dx^{\sigma}}{d\tau}=0,
\end{eqnarray}
 where  $\Gamma_{\rho\sigma}^{\mu}$ are the Christoffel connections.
Since we consider a spherically symmetrical spacetime, and after restricting the motion of atom to an equatorial plane, we take $\theta=\pi/2$, giving  $\dot{\theta}=0=\dot{\phi}$. Hence one obtains the following conservation equations,
\begin{eqnarray}
 \Big(\frac{dr}{d\tau}\Big)^2=\mathcal{E}^2-f(r),\  \Big(\frac{dr}{dt}\Big)^2=\Big(\frac{f(r)}{\mathcal{E}}\Big)^2\big[\mathcal{E}^2-f(r)\big].
\end{eqnarray}
Here, $\mathcal{E}$ represents specific energy of the atom. It is important to note that, in principle,  $\mathcal{E}^2=f(r)|_{\text{max}}$.
For asymptotically flat spacetime, $f(r)|_{\text{max}}=1$ which actually corresponds to $r\rightarrow \infty$ , hence we have 
\begin{eqnarray}
 \Big(\frac{dr}{d\tau}\Big)^2=1-f(r);\ \Big(\frac{dr}{dt}\Big)^2=f^2(r)\left[1-f(r)\right].
\end{eqnarray}
For an ingoing trajectory,  when given the initial and final positions of the atom as $r_{i}$ and $r_{f}$, respectively, the expressions  for the coordinate time $t(r)$ and the proper time $\tau(r)$ can be written as
\begin{eqnarray}\label{t}
 \tau(r)=- \int_{r_{i}}^{r_{f}} \frac{dr}{\sqrt{1-f(r)}};\ \ t(r)=-\int_{r_{i}}^{r_{f}}\frac{dr}{f(r)\sqrt{1-f(r)}}
\end{eqnarray}
Making use of Eq. (\ref{frDM}), it is possible to analytically solve the equation for  $\tau(r)$, which is expressed by 
\begin{align}
  \tau(r)&=-\sqrt{\frac{2}{3}}\alpha   \exp\left(\frac{3 M}{ \alpha }\right) \Gamma \left(\frac{1}{2},\frac{3 M}{ \alpha }-\frac{3}{2} \ln \left(\frac{r}{\alpha  }\right)\right)+\tau_{0},
\end{align}
where $\Gamma \left( \cdot \right)$ is the incomplete gamma function, and $\tau_{0}$ the constant of integration. For $t(r)$, it is difficult to obtain its general analytical expression, hence we numerically solve it to be used in Sec. \ref{secpDM}. These quantities are plotted in Fig. \ref{timeplotDM}.

The behaviours of $t(r)$ and $\tau(r)$ demonstrate the typical Schwarzschild-type character. The only difference is that PFDM increases both of them when $\alpha<\alpha_{c}$. This  should be  obvious as it takes more time to reach the shrunken BH horizon. The same reasoning goes for $\alpha>\alpha_{c}$, where both $t$ and $\tau$ decrease as the BH inflates now, and consequently the atom takes less time to cross the BH horizon.

\subsection{Field modes}
The wave equation for a massless Klein-Gordon field in the minimal coupling is given by $\nabla_{\mu}\nabla^{\mu} \Phi=0$ \cite{Birrell:1982ix}, which
for  spherical symmetry  and the timelike Killing vector $\partial_{t}$, furnishes the solution  
 $\Phi=\frac{1}{r}Y_{l}(\theta,\phi)\psi(t,r)$,
where $Y_{l}$ are spherical harmonics and $l$ is the multipole number. After neglecting $(\theta,\phi)$-dependence $(l=0)$, the radial part obeys a Schr\"odinger-type  wave equation
\begin{eqnarray}\label{WE}
 \Big(-\frac{\partial^2 }{\partial t^2}+\frac{\partial^2 }{\partial r_{*}^2}\Big)\psi(t,r)=V(r)\psi(t,r),
\end{eqnarray}
where $V(r)$ is the effective potential offered by the spacetime, and is responsible for creating scattering effects. It can be ignored in our analysis given the assumption that the emitted radiation frequency $\nu$ is large enough  to overcome this scattering \cite{Scully:2017utk,Bukhari:2022wyx}.  Thus the solution
\begin{eqnarray}\label{fieldmodeDM}
 \psi(t,r)=\exp{\left[i\nu(t- r_{*})\right]},
\end{eqnarray}
 represents an outgoing Boulware field mode detected by an observer at asymptotic distances with frequency $\nu$. The ingoing modes are lost into the boundary at the horizon.\\
 \indent 
It is pertinent to extend this discussion here by carefully considering the definition of vacuum state and exclusion of the potential. While in Minkowski spacetime, the field vacuum state is Poincar\'e invariant and hence possesses unique definition for every inertial detector. However, in curved spaces,  it becomes hard to define an \textit{objective} vacuum state \cite{Birrell:1982ix,2009qftc.book.....P}. In our case, the observations as done by the asymptotic observer really makes it plausible to assume the field to be in a Boulware vacuum state, for which normally there is no Hawking emission at asymptotic infinity.   Other possible vacuum states which are quite frequently used in these situations include Unruh  and Hartle-Hawking states \cite{Birrell:1982ix,2009qftc.book.....P}; however such states are not relevant here as the radiative analysis is done specially via the perspective of asymptotic observer for which Boulware state is the suitable state. Moreover, as we have used mirror to shroud the BH completely, it serves to shield the falling atom from any Hawking quanta, as also found in Ref. \cite{Scully:2017utk}. That said, one makes sure that the observer only receives acceleration radiation flux from the atom with no contributions from Hawking quanta.  Also, our consideration of $l=0$ modes above is the simplest case one can consider.  In fact smaller $\nu$ can be taken ($V(r)\neq 0$), where scattering effects would emerge and some interesting behavior could be expected by analyzing the situation via greybody factors \cite{Sakalli:2022xrb}.

\section{Particle Spectrum}\label{secpDM}
We first assume the field to be in Boulware vacuum state\cite{Birrell:1982ix}, so that there is no Hawking flux received by the asymptotic observer.  Neglecting angular dependence of radiation modes, the falling atom interacts with the field pictured by the Hamiltonian \cite{Scully:2017utk}
\begin{align}\label{hamiltonianDM}
 V(\tau)&=\hbar g \big[\hat{a_{\nu}}\psi(t(\tau),r(\tau))+h.c\big]\big[\hat{\sigma}(\tau)e^{-i\omega\tau}+h.c\big],
\end{align}
where $g$ is the coupling frequency that signifies strength of interaction,  $\hat{a_{\nu}}$ is the annihilation operator for the field modes,  $\hat{\sigma}$ is atomic lowering operator, and \textit{h.c} the Hermitian conjugate.  Using time-dependent perturbation theory, the excitation probability for the atom to  make a transition from a ground state $|b\rangle$ to excited state $|a\rangle$ while emitting a photon of frequency $\nu$ is given by
\begin{eqnarray}\label{exc1DM}
 P_{ex}=\frac{1}{\hbar^2}\bigg|\int d\tau\langle 1_{\nu},a|V(\tau)|0,b\rangle\bigg|^2.
\end{eqnarray}
This simultaneous atomic excitation and photon emission is rooted in acceleration, dictated by Unruh effect \cite{Unruh:1976db}.
Making use of Eq. (\ref{hamiltonianDM}), and some further calculations, one can write down Eq. (\ref{exc1DM}) as
\begin{align}\nonumber
 P_{ex}&=g^2\bigg|\int d\tau \psi^{*}(t(\tau),r(\tau))e^{i\omega\tau}\bigg|^2\\
 \label{pexx1}
 &=g^2\bigg|\int dr \bigg(\frac{d\tau}{dr}\bigg) \psi^{*}(r)e^{i\omega\tau}\bigg|^2,
  \end{align}
  which after simplification becomes 
\begin{widetext}
  \begin{align}\label{probalpha}
  P_{ex}&= g^2\Bigg|\int_{\infty}^{r_{g}} dr \exp{\left[i\nu\left\{t(r)- r_{*}(r)\right\}\right]} \frac{1}{\sqrt{\frac{2M}{r}-\frac{\alpha}{r}\ln{\left(\frac{r}{\alpha}\right)}}}
   \exp{\left[-i\omega\left\{\sqrt{\frac{2}{3}}\alpha   \exp\left(\frac{3 M}{ \alpha }\right) \Gamma \left(\frac{1}{2},\frac{3 M}{ \alpha }-\frac{3}{2} \ln \left(\frac{r}{\alpha  }\right)\right)\right\}\right]}\Bigg|^2,
\end{align}
  which can be computed numerically.  First, we consider $t$ and $r_{*}$ from Eqs. (\ref{t}) and (\ref{rstar}) for whole atomic trajectory, respectively given by
  \begin{eqnarray}\label{coordtime}
      t(r)=-\int_{\infty}^{r_{g}} \frac{dr}{\left[1-\frac{2M}{r}+\frac{\alpha}{r}\ln{\left(\frac{r}{\alpha}\right)}\right]\sqrt{\frac{2M}{r}-\frac{\alpha}{r}\ln{\left(\frac{r}{\alpha}\right)}}},\ \  r_{*}&=\int_{r_{g}}^{\infty} \frac{dr}{1-\frac{2M}{r}+\frac{\alpha}{r}\ln{\left(\frac{r}{\alpha}\right)}}.
  \end{eqnarray}
Let's make the substitution $r=r_{g}z$, such that $dr=r_{g}dz$. Hence for $t(r)$ from Eq. (\ref{coordtime}), we have 
\begin{eqnarray}
      t(z)=-\int_{\infty}^{1} \frac{r_{g}dz}{\left[1-\frac{2M}{r_{g}z}+\frac{\alpha}{r_{g}z}\ln{\left(\frac{r_{g}z}{\alpha}\right)}\right]\sqrt{\frac{2M}{r_{g}z}-\frac{\alpha}{r_{g}z}\ln{\left(\frac{r_{g}z}{\alpha}\right)}}}.
  \end{eqnarray}
We  further make substitute  $x=z-1$, such that $z=x+1$, we get 
\begin{eqnarray}\label{ttime}
      t(x)=\int_{0}^{\infty} \frac{r_{g}dx}{\left[1-\frac{2M}{r_{g}(x+1)}+\frac{\alpha}{r_{g}(x+1)}\ln{\left(\frac{r_{g}(x+1)}{\alpha}\right)}\right]\sqrt{\frac{2M}{r_{g}(x+1)}-\frac{\alpha}{r_{g}(x+1)}\ln{\left(\frac{r_{g}(x+1)}{\alpha}\right)}}}.
  \end{eqnarray}

\begin{figure*}[ht!]
\centering
\includegraphics[width=0.9\linewidth,height=13cm]{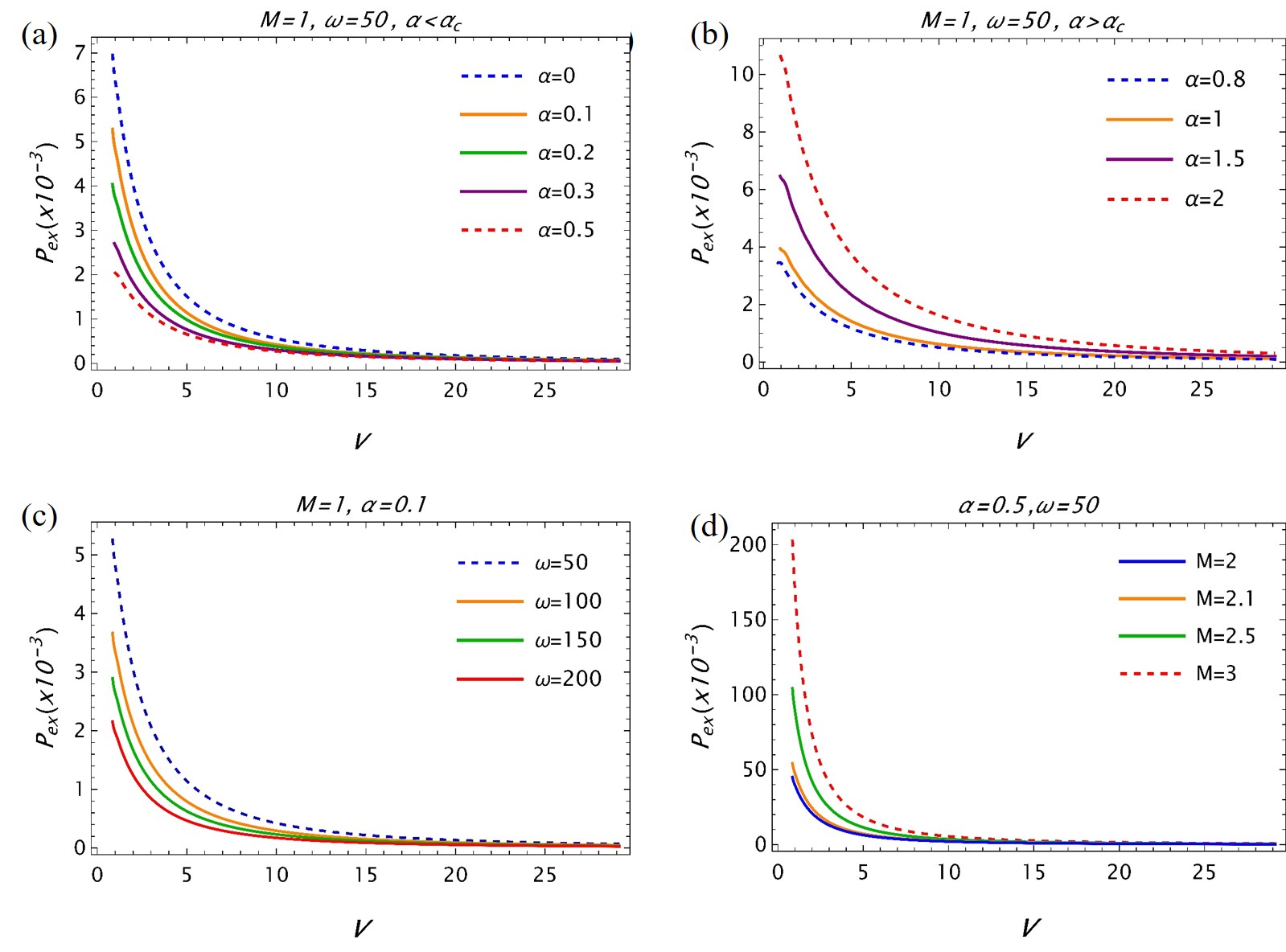}\caption{Radiation intensity $P_{ex}$ vs frequency $\nu$, affected by DM density  $\alpha$ for the regime (a) $\alpha<\alpha_{c}$, (b)  $\alpha>\alpha_{c}$, (c) atomic frequency $\omega$, and (d) BH mass $M$. The normalization of probability is done by suitable choice of $g$, which we take $g=10^{-3}$ here.}
\label{prob}%
\end{figure*}
\begin{figure*}[htb!]
\centering
\includegraphics[width=8.0cm, height=6.0cm]{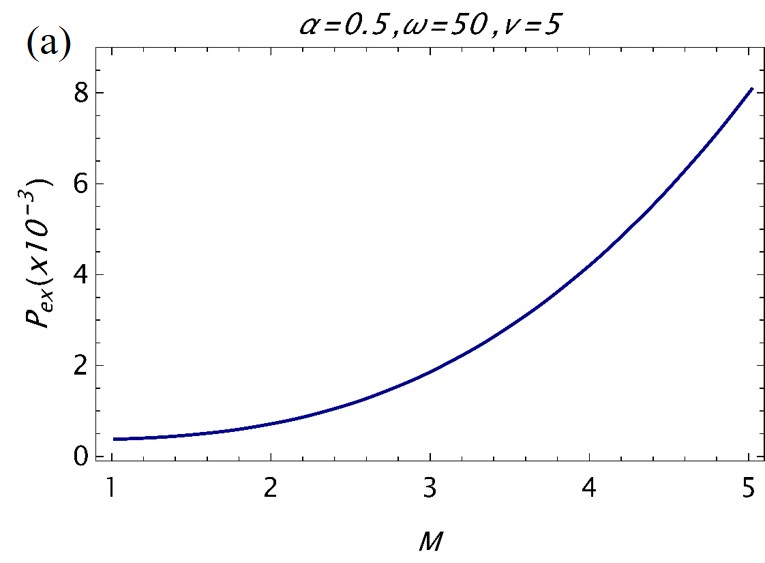}\ \ 
\includegraphics[width=8.0cm, height=6.0cm]{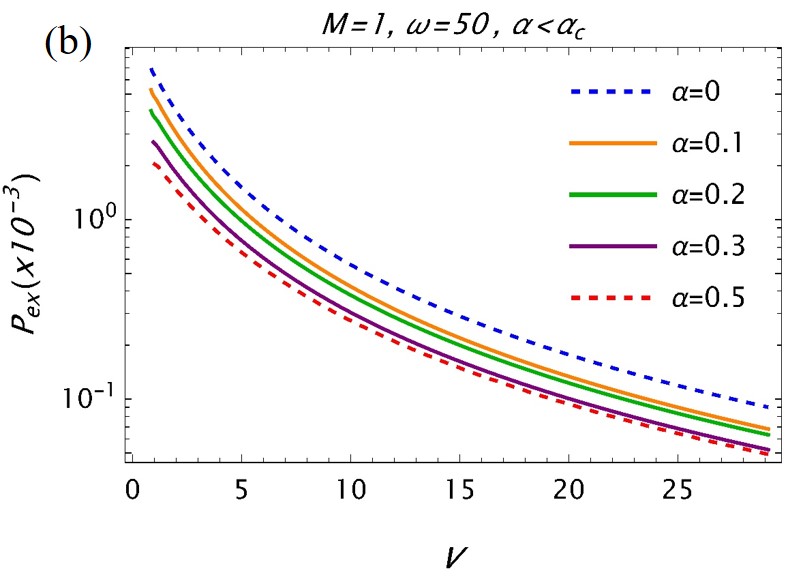}
\caption{(a) $P_{ex}$ vs BH mass $M$. $M$ enhances the particle emission nonlinearly, (b) $P_{ex}$ on a log scale.  $P_{ex}$ is a BE-type distribution being finite as $\nu\rightarrow 0$.}
\label{problog}%
\end{figure*}
Similarly, for $r_{*}$, we get
\begin{align}
    r_{*}(x)&=\int_{0}^{\infty} \frac{r_{g}dx}{1-\frac{2M}{r_{g}(x+1)}+\frac{\alpha}{r_{g}(x+1)}\ln{\left(\frac{r_{g}(x+1)}{\alpha}\right)}}.
\end{align}
Substituting all the necessary elements into Eq. (\ref{probalpha}), we get the  probability expression as follows 
\begin{align}\nonumber
  P_{ex}&= g^2r_{g}^2\left|\int_{0}^{\infty} dx \frac{\exp{\left[i\nu\left\{t(x)- r_{*}(x)\right\}\right]}} {\sqrt{\frac{2M}{r_{g}(x+1)}-\frac{\alpha}{r_{g}(x+1)}\ln{\left(\frac{r_{g}(x+1)}{\alpha}\right)}}}   \exp{\left[-i\omega\left\{\sqrt{\frac{2}{3}}\alpha   \exp\left(\frac{3 M}{ \alpha }\right) \Gamma \left(\frac{1}{2},\frac{3 M}{ \alpha }-\frac{3}{2} \ln \left(\frac{r_{g}\left(x+1\right)}{\alpha  }\right)\right)\right\}\right]}\right|^2,
\end{align}
\end{widetext}
which holds the crux of our work. We now numerically solve it to see the effects of PFDM and other parameters. The plots are given in Figs. \ref{prob} and \ref{problog}.

\section{Results and discussions}\label{discussionDM}
Keeping the above formulation in view, it is intriguing to have the emission of radiation even though the field is in a Boulware vacuum state for which normally there is no Hawking flux at asymptotic infinity. However, this particular emission has stark contrasts to that of Hawking radiation. Here, the field evolves as a pure state and possesses phase correlations \cite{Scully:2017utk}. One can also trace the origins of this radiation to near horizon approximation and conformal quantum mechanics \cite{Camblong:2020pme, PhysRevD.104.084086,Azizi:2021yto}.   \\
\indent The above plots contain the impact on excitation probability $P_{ex}$ by PFDM density $\alpha$ as shown in Figs. \ref{prob}(a) and \ref{prob}(b), transition frequency of the atom $\omega$ in Fig. \ref{prob}(c), and BH mass $M$ in Fig. \ref{prob}(d). We attempt to  encapsulate the underlying physics in the following way.\\
\indent In pure BHs with asymptotically flat spacetime, earlier studies \cite{Scully:2017utk,PhysRevD.104.084086,Chakraborty:2019ltu} have reported thermal generation of particles depicting either a Planckian or a Bose-Einstein (BE) distribution.  As our plots also show a thermal BE-type distribution for field quanta, it seems reasonable to conclude that PFDM keeps the thermality intact. Although thermality is not broken, the excitation probability shows two distinct behaviors depending on whether the PFDM density $\alpha$ is less or greater than the critical density $\alpha_{c}$ given by the Eq. (\ref{alphac}).   \\
\indent When the PFDM density is less than the critical density, i.e. $\alpha<\alpha_{c}$, the probability diminishes as we approach $\alpha_{c}$. In particular, for the choice for BH mass $M=1$, $\alpha_{c}\sim 0.538$ as noted earlier. The amplitude is highest for the least value of $\alpha$, i.e. $\alpha=0.1$, shown as solid orange line in Fig.\ref{prob} (a).The decreasing trend for amplitude continues till the largest $\alpha$, i.e. $\alpha=0.5$, depicted by dashed red curve  in Fig. \ref{prob} (a). We note a very crucial point here.  Compared to a pure Schwarzschild case $(\alpha=0)$, shown by dashed blue curve in Fig.\ref{prob} (a),  BHs surrounded with PFDM will always have lesser particle creation profile provided one ensures $\alpha<\alpha_{c}$.\\
\indent In a stark contrast, once we cross $\alpha=\alpha_{c}$, the probability amplitude increases as seen from  Fig. \ref{prob} (b), while remembering the maximum value allowed for $\alpha=2M$. The overall situation may be somewhat  intuitively pictured as follows.\\
\indent  There is a strong evidence in Hawking's scenario that particles once leaving the BH horizon suffer from backreaction from  tidal forces of the BH \cite{Dey:2017yez}, which weakens the strength of flux.  Tidal forces originate directly out of surface gravity of the BH, and surface gravity stems out of the size of the event horizon of the BH. Quantitatively speaking, for a BH with a given horizon radius $r_{g}$, surface gravity  $\propto 1/r_{g}^2$, which obviously translates to the fact that smaller BHs have large surface gravity, i.e., tidal forces, while supermassive BHs have negligible surface gravity. In the present case, for the PFDM regime characterized by $\alpha<\alpha_{c}$, the BH radius consistently decreases thereby increasing the surface gravity  of the BH and hence the tidal forces. This increasing surface gravity compared to the pure Schwarzschild BH $(\alpha=0)$  forces radiation flux to experience more backreaction from the geometry, resulting in a decreased intensity, as obvious from Fig. \ref{prob} (a).  \\
\indent Conversely, for the PFDM regime $\alpha>\alpha_{c}$, the BH size increases continuously for all larger values of $\alpha$ (lesser surface gravity), by virtue of which our spacetime offers lesser backreaction. Hence the enhancement of flux.  This situation is again witnessed for the case when BH mass $M$ increases [see Fig.\ref{prob} (d)], and is grounded in the same reasons as stated above. To get  a more clear insight into the role of $M$, we plot $P_{ex}$ against $M$ in Fig.\ref{problog}(a). We see that $M$ monotonically increases $P_{ex}$ in a nonlinear fashion. One infers from this that the particle spectrum for supermassive BHs immersed in PFDM distribution should be richer than the stellar or intermediate-mass BHs.\\
\indent By realizing the fact that PFDM seems to make its impact by merely  rescaling the BH size, it might be tempting to conclude the close similarity the situation may bear with. For example,  there is a similar geometric structure for a  Reissner-Nordstr\"om BH  with the same mass--at least for the range $\alpha<\alpha_{c}$--one could well argue that it might also generate similar particle spectrum to that of Figs. \ref{prob} and \ref{problog}. We however note that there is a considerable difference between BHs with hair (like Reissner-Nordstr\"om BH) which usually possess multiple horizons and BHs with PFDM halo (no hair). The role of charge on Ressiner-Nordstr\"om BH in Hawking radiation and quasinormal (QNM) spectrum is quite well known \cite{Jiang:2005xb,Destounis:2021lum}. In fact, Reissner-Norstr\"om BHs has distinct features which are absent in pure Schwarzschild BH, such as existence of long-lived QNM modes near extremal case where charge and mass balance each other. In contrast to this, our BH is geometry is very much Schwarzschild-like and would be expected to possess corresponding particle spectrum profile.\\
\indent We also attempted to quantify the role of atomic transition frequency $\omega$ on radiation flux as shown in Fig.\ref{prob} (c). Any increment in $\omega$ would make it difficult to excite an atom and thus degrades the emission. This corroborates to the standard Unruh effect \cite{Unruh:1976db}. In passing, to check whether BE-type distribution diverges or stays finite near the origin $(\nu\rightarrow 0)$, we plotted $P_{ex}$  on log-scale as shown in Fig. \ref{problog}(b).
It is evident that the spectrum remains finite at $\nu\rightarrow 0$. We also declare here that all numerical computations  were performed in \textit{Mathematica} software.\\
\indent Meanwhile, one may wonder whether there is any possibility of having testable aspects of such study. To that end, we may state the following. We do not claim that this work may entail any direct observational/experimental  consequences for DM research. However, it is encouraging to seek the possible relevance to  analogue gravity systems \cite{Barcelo:2005fc,Braunstein:2023jpo,Jacquet:2020bar}. Analogue gravity experiments have recently been very promising in providing rich insights into the trapped horizons and the associated phenomena, such as Hawking radiation \cite{Weinfurtner:2010nu}, Unruh effect \cite{Hu:2018psq}, and expanding spacetimes  involving cosmological particle creation \cite{Steinhauer:2021fhb}. It is tempting to note that these cosmological phenomena bear same underlying principle to that of  dynamical Casimir effect, as mentioned in the beginning of this work. Most of the analogue gravity experiments are realized by manipulating condensed matter and optical systems. In particular, condensed matter systems are becoming popular test beds for phenomena beyond quantum field theoretic effects, such as fluid/gravity correspondence \cite{Hubeny:2011hd}.  Lets also note that DM research is being increasingly devoted to getting any signatures via condensed matter systems \cite{Kahn:2021ttr,Prabhu:2022dtm}. Gathering these facts that cosmological processes can be realized in quantum matter and optics, it seems conceivable to predict that future table-top experiments might also possibly exploit these systems to simulate dark matter candidates. The same logic goes for this particular acceleration radiation from atoms in free fall \cite{Scully:2017utk}.   It may be true to state that one should not expect any immediate advances for such avenues as of now. We nevertheless could in principle expect such possibilities in far future, however daunting it may be.  It is in this line that the particle spectrum profile associated with PFDM reported here may perhaps seem to make any relevance.\\
\indent Finally, we remark that we attempted to provide the most plausible explanation for our findings on physical grounds. Hence, our results should not be taken as mere mathematical artifact.      
\setlength{\parskip}{0cm}
    \setlength{\parindent}{1em}
\section{Conclusive remarks}\label{secconclusions}
\setlength{\parskip}{0cm}
    \setlength{\parindent}{1em}
Dark Matter (DM) is the mysterious invisible stuff that is believed to be permeating all galaxies, dictating the structure formation in the Universe. Even though it is a powerful model to explain a plethora of observational aspects in the Universe, it nevertheless does not surface in direct experimental setups. Modelling  and searching for DM is one the pressing problems in modern cosmology and astrophysics. Among these models, perfect fluid dark matter (PFDM) has been one of potential hot candidates in recent times. \\ 
\indent Given that precision techniques in DM search have made great progress recently, the central concern about DM then is whether it interacts with ordinary matter, and what and how it leaves its imprints. It thus becomes increasingly important to look for DM signatures in all possible known phenomena.\\
\indent Our work explores one such avenue by relating DM to the Casimir paradigm. We analyzed the acceleration radiation emitted by an atom interacting with a Casimir boundary held fixed at the event horizon of a Schwarzscihld BH surrounded by a PFDM halo. We showed that PFDM can decrease or increase the intensity of emitted quanta with sole dependence on the its density.  This degrading occurs if PFDM density stays below a certain critical value, and vice versa. The impact of atomic transition frequency $\omega$ and BH mass $M$ were also quantified. \\
\indent Our work can be generalized to other models of DM, in either classical or quantum gravity domain. In addition to this, one can go beyond this simplistic quantum optical model to invoke other formulations of atom-field dynamics. \\
\indent Admitting the fact that our setup is a Gedanken experiment, we however believe there still is a great scope for deciphering the nature of PFDM physics. Hence this study may possibly provide any hints in constraining DM parameter space guided by precision experiments in the future.     
\acknowledgments{ This research is supported by the National Natural Science Foundation of China (NSFC) (grant No. 11974309)}. SMASB acknowledges financial support from China Scholarship Council at Zhejiang University.
\bibliographystyle{apsrev4-1}
\bibliography{masood.bib}
\end{document}